# Grain boundary segregation in steels: Towards engineering the design of internal interfaces

Mainak Saha

Department of Metallurgical and Materials Engineering, Indian Institute of Technology Madras, Chennai-600036, India

Corresponding authors: Mainak Saha

Email address (es): mainaksaha1995@gmail.com

Phone number: +918017457062

ORCID: Mainak Saha: 0000-0001-8979-457x

**Abstract**

Solute decoration at grain boundaries (GB) leads to a number of phenomenon such as changes in interface structure, mobility, cohesion etc. Recent experimental investigations on interfacial segregation in steels are based on microstructural characterisation using two correlative methodologies, namely, Transmission Electron Microscopy-Atom Probe Tomography (APT) and Electron Backscatter Diffraction-APT. Considering the growing interest in this avenue, the present review is aimed at addressing the common adsorption isotherms used for quantifying interfacial segregation and providing an overview of the present state of experimental research in the area of GB segregation in steels. The areas where an understanding of GB segregation may be utilised have also been highlighted with a focus on the experimental challenges associated with understanding GB segregation in steels.

**Table of contents**

**1. Introduction**

In the context of polycrystalline metallic materials, grain boundaries (GBs) are planar (2D) defects which influence a wide range of properties such as tensile strength. fatigue resistance, thermal and electrical conductivity, corrosion, resistance to hydrogen (H) attack etc. [1]–[5]. Being a region of partial atomic disorder but with definite structure and orientation, a GB may also act as a source and sink for vacancies and dislocations [6]–[9]. In addition, they possess five macroscopic degrees of freedom (DOFs) (which include misorientation angle and axis) [10]–[15]. The five DOFs determine the structure of a GB [16]–[19]. The energy of a GB (also known as GB energy) is highly sensitive to GB structure and is influenced by the local chemistry near GBs [17], [18], [20]–[24]. The driving force for solute decoration at GBs is the minimisation of GB energy [25]–[27].

Special high-angle GBs (SHAGBs) with low GB energies have been reported to possess much higher resistance to corrosion, crack propagation, H diffusion, GB sliding etc. as compared to that in random HAGBs (RHAGBs) (with high GB energies) [28]–[31]. Such observations have led to the evolution of GB engineering (GBE) which is based on replacing RHAGBs with SHAGBs for the purpose of optimising the properties (of polycrystalline materials) [32]–[37]. However, preferential segregation of different elements (with different concentration) at GBs and its associated influence needs to be considered [2], [38]–[42]. This is highly important since most of the GB properties are influenced by segregation-induced changes which mainly include fracture toughness, electrical and thermal properties, H embrittlement, resistance to dislocation pile-up etc. [2], [43]. Such manipulation of GB structure has been referred to as GB segregation engineering (GBSE) [2]. Both thermodynamic (such as co-segregation, coefficient of segregation etc.) and kinetic factors (such as deformation-induced GB phase evolution) are associated with GBSE [2], [44]–[46]. This implies that in addition to thermomechanical treatment, solute decoration at GBs is hugely influenced by time [2]. Besides, GBSE utilises elemental segregation as a methodology for site-specific manipulation which leads to the optimisation of specific GB structure, composition and properties [2].

Moreover, GBs may either reduce/increase the strength of polycrystalline metallic materials [1]. Most metallic materials, during service, tend to fail from an intergranular fracture through void nucleation and growth owing to high stress concentration along GBs [47]. Predominant GB embrittlement due to non-accommodation of plastic strains along GBs in presence of segregating solute species has been most commonly attributed to be the cause of such a failure [48], [49]. Such an embrittlement assisted failure leads to "impaired plasticity" in metallic materials [50]. A very common example of such a phenomenon is the embrittlement caused by the segregation of H at different GBs in steels [2], [51]. On the contrary, there have also been a number of reports on the GB strengthening (in metallic materials) caused due to the accommodation of plastic strains along GBs, especially in the presence of solute segregation [40], [52]–[54]. Hence, the overall aim is to design metallic materials with high overall toughness. In the context of steels, which form the backbone of the global economy, the former statement tends to be highly applicable. In other words, solute decoration at internal interfaces may be used as a design tool for developing high-performance steels [2]. The present review article is aimed at addressing the common thermodynamic approaches for quantifying the extent of interfacial segregation and highlighting the present state of experimental research in the avenue of GB segregation in steels. The avenues where an understanding of GB segregation may be utilised have also been highlighted. In addition, the experimental challenges associated with GB research have been briefly discussed from the viewpoint of solute segregation.

## 2. Theoretical approaches towards quantifying interfacial segregation

Solute concentration at GB has been reported to exceed the solubility (of solutes) in the grain interior, upto several orders of magnitude [1], [2], [48], [55]–[57]. Raabe et al. [2] have reported that bulk solubility (solubility of a particular solute within the interior of a grain) may be used as an approximation for determining the tendency of a solute towards preferential segregation at GBs. In other words, the smaller the bulk solubility, the higher is the solute enrichment factor ($\beta_i$) of the solute at the GB which implies a higher tendency of the solute species towards GB segregation [2]. **Fig. 1** shows the variation of $\beta_i$ as a function of the solubility limit of different solutes in α-Fe matrix.

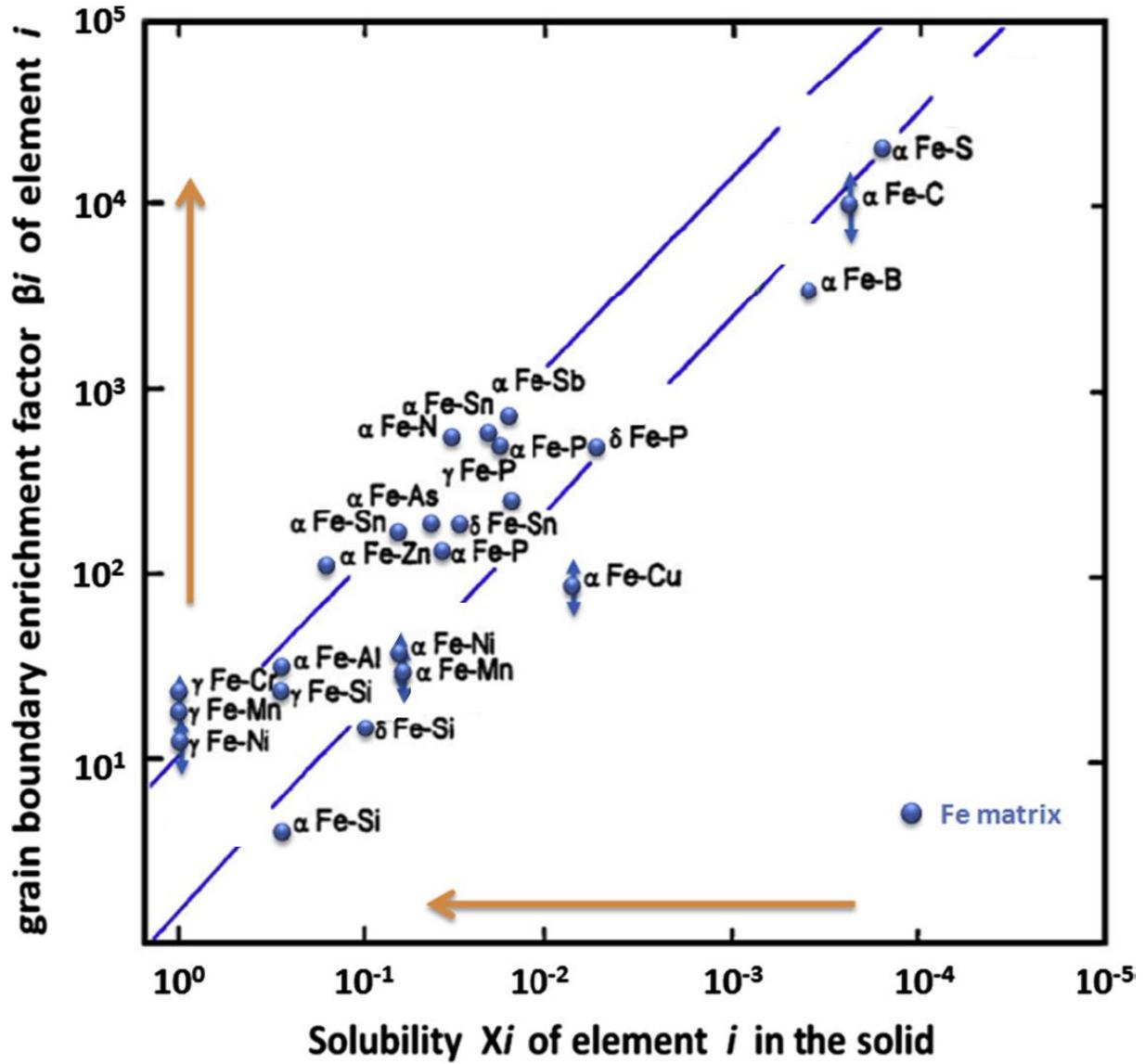

**Fig. 1** Variation of $\beta_i$ as a function of solubility limit of different solutes in Fe matrix, obtained from Ref. [2]. The data points in this plot may be used to identify appropriate solute species with a strong tendency towards GB segregation.

In addition, the thermodynamics associated with GB segregation has a close analogy to monolayer gas adsorption at solid surfaces and may be formulated in terms of the Gibbs adsorption isotherm [16], [58]. Using this isotherm, the excess concentration of a given solute species at a GB (also known as GB excess) is given as: [16]

$$\Gamma_i = -\left(\frac{1}{RT}\right)\left(\frac{d\gamma_{GB}}{dx_i}\right)_{V,T} \qquad (1)$$

where, $x_i$ are the molar concentration of the element (solute) i in the bulk (or, grain interior), $d\gamma_{GB}$ is the change in GB energy upon segregation at a fixed temperature T and volume V, R is the universal gas constant (~ 8.314 J/molK), and T is the absolute temperature (in K) [16]. Theoretically, $\Gamma_i$ may be determined by measuring $d\gamma_{GB}$ as a function of $dx_i$ in logarithmic scale. In the context of GB segregation, the other form of Gibbs adsorption isotherm is commonly used which relates $d\gamma_{GB}$ with $\Gamma_i$ and change in chemical potential of solute i (dμ$_i$). It is expressed as: [58]

$$d\gamma_{GB} = -\sum_i \Gamma_i d\mu_i \qquad (2)$$

From a thermodynamic viewpoint, Gibbs adsorption isotherm is based on a simplistic approach towards quantifying GB segregation. However, the measurement of GB segregation as a function of bulk concentration and temperature is not possible using the aforementioned isotherm [59], [60]. In this context, the Langmuir-McLean isotherm has been widely used since it quantifies GB segregation by balancing both the adsorption and desorption rates of solute species at a GB [60]. This isotherm is given as: [60]

$$\frac{x_i^{GB}}{(x_i^{GB,0} - x_i^{GB})} = \frac{x_i^B}{(1 - x_i^B)} \exp\left(-\frac{\Delta G_i^{GB}}{RT}\right) \qquad (3)$$

where, $x_i^{GB}$ is the molar concentration of solute i at GB, $x_i^{GB,0}$ is the solubility limit (in terms of molar concentration) of solute i at GB, $x_i^B$ is the molar fraction of solute i at grain interior (bulk), and $\Delta G_i^{GB}$ is the Gibbs molar free energy of segregation of solute i at GB [60]. Considering the system as dilute, equation (3) may be simplified as follows: [60]

$$\beta_i = \frac{x_i^{GB}}{x_i^B} = \exp\left(-\frac{\Delta G_i^{GB}}{RT}\right) \qquad (4)$$

$\Delta G_i^{GB}$ is usually an unknown quantity. Based on equations (3) and (4), it may be implied that GB segregation occurs when $\Delta G_i^{GB} < 0$ and with increasing temperature, there is a decrease in the tendency of segregation of solute species at GBs [60]. The major limitation of Langmuir-McLean isotherm is that it does not consider GB energy for quantification of GB segregation unlike Gibbs adsorption isotherm [2], [61]. Moreover, this isotherm assumes that dynamic equilibrium is established between the segregating solute species (in case of multiple decorating solutes) and that solute decoration (at GBs) is limited to only a single monolayer [60]. However, both the aforementioned isotherms treat GBs as an individual phase with properties which are different as compared to those in the grain interior and do not consider the interaction between segregating solute species at GBs [2].

Recently, the phenomenon of entropy-dominated GB segregation has been introduced for low solute concentration at different SHAGBs and RHAGBs in α-Fe based alloys [62]. It has been reported that the main limitation of the Langmuir-McLean isotherm is associated with the assumption that the phenomenon of GB segregation is only controlled by enthalpy of segregation ($\Delta H_i$, for solute i) and not by the entropy of segregation ($\Delta S_i$) [62], [63]. Besides, it has been theoretically shown that entropy-dominated segregation phenomenon is dominant for solutes with low solid solubility in a given matrix (solvent) especially at homologous temperatures $> 0.4 T_m$ ($T_m$: melting points (in K)) [62], [63]. In the context of α-Fe, the two common segregating species with low solid solubility are Sb and Sn [62], [64]. These elements have been reported to exhibit dominant entropy-dominated GB segregation phenomenon in α-Fe at temperatures $\geq 450°C$ [62].

Based on GB segregation maps at different temperatures, Lejček and Hoffmann [65]–[67] have shown that solubility limits for different solute atoms at different GBs varies with GB character in α-Fe. Hence, it may be argued that GB segregation diagrams provide an understanding of solute segregation in dilute binary systems. However, from a practical viewpoint, it is always more useful to know about GB segregation in multicomponent alloys where mutual interactions and site competition both simultaneously play an important role, in determining the extent of solute segregation at different GBs [50], [67]–[70]. In such cases, $\beta_i$ fails to correctly predict the amount of solute segregation at GBs [65], [68], [71]–[77]. According to Guttmann [78]–[80], the **$\Delta H_i$** for a solute i in a ternary system with matrix M and the other solute j is given by:

$$\Delta H_i = \Delta H_i^0 - 2\alpha_{Mi}(x_i^{GB} - x_i) + \alpha_{ij}(x_j^{GB} - x_j) \qquad (5)$$

where $x_i^{GB}$, $x_j^{GB}$, and $x_i$, $x_j$ are the GB and bulk mole fractions, respectively, of two solutes i and j. $\Delta H_i$ depends on the chemical interaction between the elements, i.e., the binary interaction with the Fowler interaction coefficient $\alpha_{Mi}$ and the ternary interaction coefficient $\alpha_{ij}'$ with the other solute species [81]. $\Delta H_i^0$ denotes the enthalpy of segregation for the dilute binary system M-i and is independent of temperature and composition even for a multicomponent system [80]. Therefore, $\Delta H_i^0$ in equation (5), is an independent characteristic of the segregation of solute i in matrix M [81]. Hence, it may be justified that the tendency of a solute to segregate at GBs is better characterized by a more fundamental thermodynamic quantity such as $\Delta H_i$ as compared to $\beta_i$.

## 3. Characterising GB segregation in steels

## 3.1 Common techniques for the characterisation of GB segregation

A number of characterisation techniques have been used to investigate GB segregation such as Auger Electron Spectroscopy (AES), Transmission Electron Microscopy (TEM), Scanning Transmission Electron Microscopy (STEM), Electron Backscatter Diffraction (EBSD), Transmission Kikuchi Diffraction (TKD), Electron Energy Loss Spectroscopy (EELS), Field Ion Microscopy (FIM), Secondary Ion Mass Spectroscopy (SIMS) and Atom Probe Tomography (APT). All these techniques, when used individually for characterising GB segregation, have a number of limitations which primarily include a lack of crystallographic details and a lack of GB segregation information with high spatial resolution [2]. Moreover, GB segregation is an atomic-scale phenomenon and hence, it is challenging to obtain GB segregation information which is precise, statistically-significant and reproducible at the same time [2]. Wynblatt and Chatain [82] have highlighted that there recently exists a discrepancy between the present theoretical and experimental approach towards addressing GB segregation.

One of the recently evolved methodologies towards characterising GB segregation is correlative microscopy which enables a simultaneous determination of both structural and chemical information from a region near a solute-decorated GB. The two correlative characterisation methodologies employed (till date) for the determination of GB segregation in steels are based on correlative TEM-APT and EBSD-APT approach which are based on a direct lattice reconstruction from a given APT data of solute-decorated GBs [2], [83]–[85]. Moreover, the GB segregation information obtained using electron microscopy techniques (such as TEM and EBSD) is influenced by a number of factors. First, the angular resolution of the electron microscopy technique used. For instance, using TEM-based nanobeam diffraction technique, it is possible to obtain an angular resolution of ~1° in contrast to that (~1-3°) obtained using EBSD technique, depending on the instrument and analysis software used [86]. Second, there is a huge uncertainty associated with preparing APT tips (containing GB region) using Focussed Ion Beam (FIB)-based liftout technique in a Scanning Electron Microscope (SEM) [2], [87]. Earlier, this uncertainty was addressed by conducting orientation mapping on the APT tips by using TEM and/or FIM techniques. In addition to the limited field of view, a high level of expertise is required to obtain precise orientation information (from APT tips) using these techniques. In recent times, the aforementioned uncertainty of obtaining GB probed region (using FIB technique) is eliminated by conducting orientation measurement using TKD technique and subjecting the same tips to APT analysis. In contrast to TEM and FIM techniques, TKD technique has a higher field of view [88], [89]. Besides, obtaining orientation information using TKD is much easier as compared to that using TEM and/or FIM techniques [90]. One of the limitations of APT is the influence of projection and lens effects on the allocation of atoms to a particular GB [91]–[93]. The projection effect is based on the magnetic field surrounding the APT tip whereas the lens effect leads to errors in the positioning of field evaporated atoms and is associated with field evaporation at solute-decorated GBs [91].

**3.2 Experimental investigations towards understanding GB segregation in steels**

**3.2.1 Conventional methodology**

Grabke et al. [94] have used FIM, APT and AES techniques to investigate the segregation of Ti, Nb, Mo and V and their carbides at RHAGBs in Fe-P based ternary and quaternary alloys (bulk composition indicated in Ref. [94]). Addition of Ti and Nb was reported to prevent the segregation of P at GBs (through formation of Ti and Nb-based phosphides) in Fe-Ti-P ternary alloys [94]. However, both Mo and V were reported not to influence the segregation tendency of P [94]. In the context of Fe-Nb-C-P and Fe-Ti-C-P quaternary alloys, GB segregation of P was reported to increase with C concentration upto a point at which NbC precipitates form [94]. Using AES, Christien et al. [95] have established a linear relationship between intergranular grooves formed during metallographic etching with the concentration of P segregated at GBs in a 17-4PH martensitic stainless steel. Takahashi et al. [96] have investigated the segregation of C and N atoms at different GBs in Fe-0.006C-0.001N-0.04Al (C60) and Fe-0.005C-0.0054N-0.004Al (N60) (wt.%) ferritic steels using APT. Based on the observations, it was highlighted that C segregates at RHAGBs in C60 whereas N segregates at RHAGBs in N60 [96]. However, the segregation tendency of N in N60 was determined to be lesser than that of C in C60 [96]. In addition, the Hall-Petch coefficient of ferrite grains was reported to be influenced by the segregation of C atoms at RHAGBs (in C60) and not by the segregation of N atoms at RHAGBs (in N60) [96].

Rosa et al. [97] have investigated the influence of B addition to Fe-0.34C-2.45Mn-0.0100B-0.03Ti (at.%) high strength steel (with a single-phase austenitic structure) using nano-SIMS and APT techniques. Dissolution of boride ($Fe_2B$ and $M_{23}(B,C)_6$) precipitates at grain interior leads to segregation of B at RHAGB [97]. Besides, the extent of B segregation at austenite GBs was reported to increase with temperature [97]. Although this goes against the thermodynamically supported trend of decreasing solute segregation with increasing temperature, however, B segregation at austenite GBs was reported to follow Langmuir-McLean adsorption isotherm [97]. Based on this isotherm, the $\Delta H_i$ of B (at austenite GB) and the enthalpy of dissolution ($\Delta H_{diss}$) of borides were calculated to be higher (by ~51% and 9.4%) than the respective values (of $\Delta H_i$ and $\Delta H_{diss}$) reported in the earlier literatures [98]–[100]. Fedotova et al. [101] have investigated irradiation-induced GB segregation of P, Mn, Ni and Si at RHAGBs in Fe-(0.04-0.07)C-(1.6-1.89)Ni-(0.006-0.01)P (wt.%) reactor pressure vessel steels using APT, AES and fractography techniques. The extent of segregation of these elements (at RHAGBs) was reported to increase with increasing bulk concentration [101]. Besides, segregation of P at RHAGB was observed to promote inter-granular fracture [101]. A similar tendency of P has been reported in Fe-(0.083-0.19)C-(0.26-0.38)Si-(1.22-1.41)Mn-(0.007-0.012)P (wt.%) reactor pressure vessel steels [102] and in neutron-irradiated AISI 304 austenitic stainless steel [103], [104].

Shigesato et al. [105] have measured non-equilibrium segregation of B at austenite RHAGBs in Fe-0.05C–0.5Mo–0.001B (wt.%) steel using aberration-corrected STEM-EELS technique. Moreover, the influence of specimen thickness, electron beam broadening, and GB plane inclination on the GB segregation of B was also investigated in this work [105]. Gaussian broadening model was used to determine the broadening of B concentration profile [105]. It was reported that the extent of broadening increases with specimen thickness [105]. However, for specimen thickness < 30 nm, the broadening of B concentration profile was obtained to be < 10% [105]. Besides, B concentration profile was also reported to be asymmetric for GB plane inclination angle > 1.5° [105]. Ma et al. [106] have used EBSD and APT techniques to highlight that segregation of C at ferrite/austenite interphase boundary (IB) leads to an increase in the energy barrier for dislocation emission from RHAGBs in a duplex medium Mn steel (composition: Fe-11.7Mn-2.9Al-0.064C (wt.%)) resulting in discontinuous yielding of the material at room temperature. C-decoration at the aforementioned IB coupled with dislocation nucleation and subsequent multiplication has been reported to influence the discontinuous yielding phenomenon [106]. Besides, it was also reported that segregation of solutes at IBs is not governed by Gibbs and Langmuir-McLean adsorption isotherms [106] unlike the previous reports [107]–[109].

### 3.2.2 Correlative methodology

**TEM-APT methodology**

Herbig et al. [86] have characterised C-decorated GBs in a nanocrystalline cold-drawn pearlitic steel (composition: Fe–4.40C–0.30Mn–0.39Si–0.21Cr (at. %)) using correlative TEM-APT methodology and reported the variation of C excess (obtained using APT) as a function of GB misorientation angle for coherent Σ5 and both coherent and incoherent Σ3 GBs (**Fig. 2**). **Fig. 2(a)** shows the overlay of C decoration at different GBs in an APT needle. **Fig. 2(b)** shows the variation of C excess as a function of GB misorientation angle. For incoherent Σ3 GBs, C-excess was found to be much higher as compared to that of coherent Σ5 and Σ3 GBs [86]. Besides, the deviation from the ideal 60° misorientation (for incoherent Σ3 GBs) was attributed to the presence of misfit dislocations which, in addition, are also responsible for high C-excess at the aforementioned GB (**Fig. 2(b)**) [86]. Moreover, the C excess vs GB misorientation angle plot was found to possess a good correlation with GB energy vs misorientation angle plot for the aforementioned GBs [86]. Abramova et al. [110] have used the same methodology to highlight that segregation of Mo, Si and Cr at austenitic RHAGB leads to GB strengthening in ultra-fine grained AISI 316 austenitic stainless steel. In the context of Fe-1.62Mn-0.18Si (wt.%) ferritic steel, Han et al. [111] have used the aforementioned methodology to report that delamination is highly influenced of competitive segregation of C and P (at RHAGBs). It was also highlighted that delamination cracks are promoted along RHAGBs with high P and low C content whereas RHAGBs with low P and high C content were observed to be resistant to delamination cracking [111].

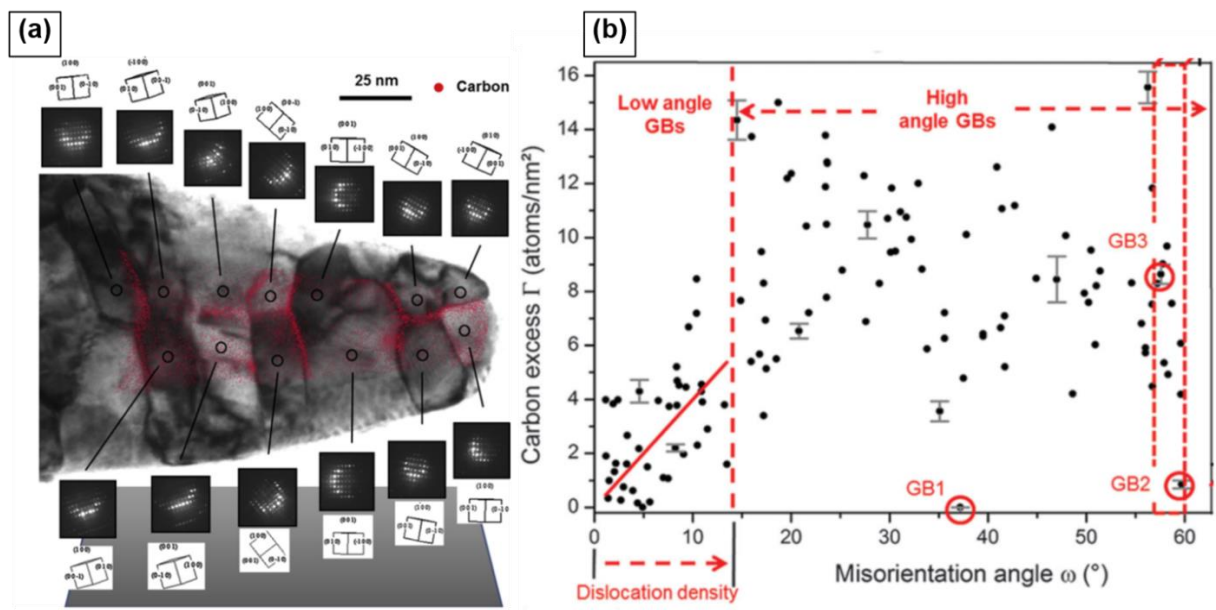

**Fig. 2** Correlative TEM-APT analysis for Fe–4.40C–0.30Mn–0.39Si–0.21Cr (at. %) cold drawn pearlitic steel: **(a)** Overlay of C decoration at different GBs in an APT needle prepared using FIB-based liftout technique and **(b)** Variation of C excess as a function of GB misorientation angle [86].

**EBSD-APT methodology**

Kuzmina et al. [112] have used this methodology to highlight that Mn segregation leads to the embrittlement of martensite RHAGBs in Fe-9Mn-0.05C (wt.%) steel at 450°C. In that context, addition of ~ 0.0027 wt.% of B was reported to promote GB strengthening and prevent the segregation of Mn at martensite RHAGB by promoting martensite to austenite reversion after prolonged holding (for ~ 336h) at 450°C [112]. A similar observation was reported in the context of Fe-12Mn-3Al-0.05C (wt.%) steel by Benzing et al. [113]. Ravi et al. [114] have utilised this methodology to study the influence of C segregation in Fe–0.2C–3Mn–2Si (wt.%) bainitic steel and reported that segregation of C at austenite RHAGBs promotes austenitic to bainitic transformation during isothermal holding at bainitic transformation temperature (~400°C, in this case).

Using the same approach, Herbig et al. [115] have characterised the segregation of B, C, P, Si and Cu at an FCC RHAGB and coherent $\Sigma 3$ annealing twin boundaries in Fe–28Mn–0.3C (wt.%) twinning-induced plasticity (TWIP) steel (**Fig. 3**). **Fig. 3(a)** shows EBSD-based image quality (IQ)+GB map. **Figs. 3(b and c)** show the APT-based 3D concentration profiles of B, C, Si, P, Mn and Cu at the two GBs highlighted (using black outlined rectangular boxes) in **Fig. 3(a)**. The extent of Mn segregation was observed to be nearly similar for both RHAGB and at $\Sigma 3$ GB (**Figs. 3(b and c)**)). B and P were reported to segregate much strongly at RHAGB than at coherent $\Sigma 3$ annealing twin boundaries (**Figs. 3(b and c)**) [115]. This was attributed to the higher GB energy of RHAGB as compared to that of coherent $\Sigma 3$ GB [115]. Decoration of RHAGB by C atoms (**Fig. 3(b)**) was reported to increase the local SFE and subsequently, lead to high resolved shear stress for mechanical twinning (at RHAGB) [115]. On the contrary, depletion of C (at $\Sigma 3$ GB (**Fig. 3(c)**)) was reported to lower the local SFE leading to the formation of ε-martensite (HCP) phase at austenite $\Sigma 3$ boundaries [115]. Besides, the preferential segregation of elements (B, C, P and Si) at deformation twins in a deformed Fe–22Mn–0.6C (wt.%) TWIP steel was investigated [115]. It was reported that although crystallographically similar, deformation twins show a much lower tendency of undergoing decoration by C atoms as compared to that of annealing twins [115]. This tendency of deformation twins may be attributed to the extremely low mobility of C atoms during the formation of deformation twins at room temperature [115].

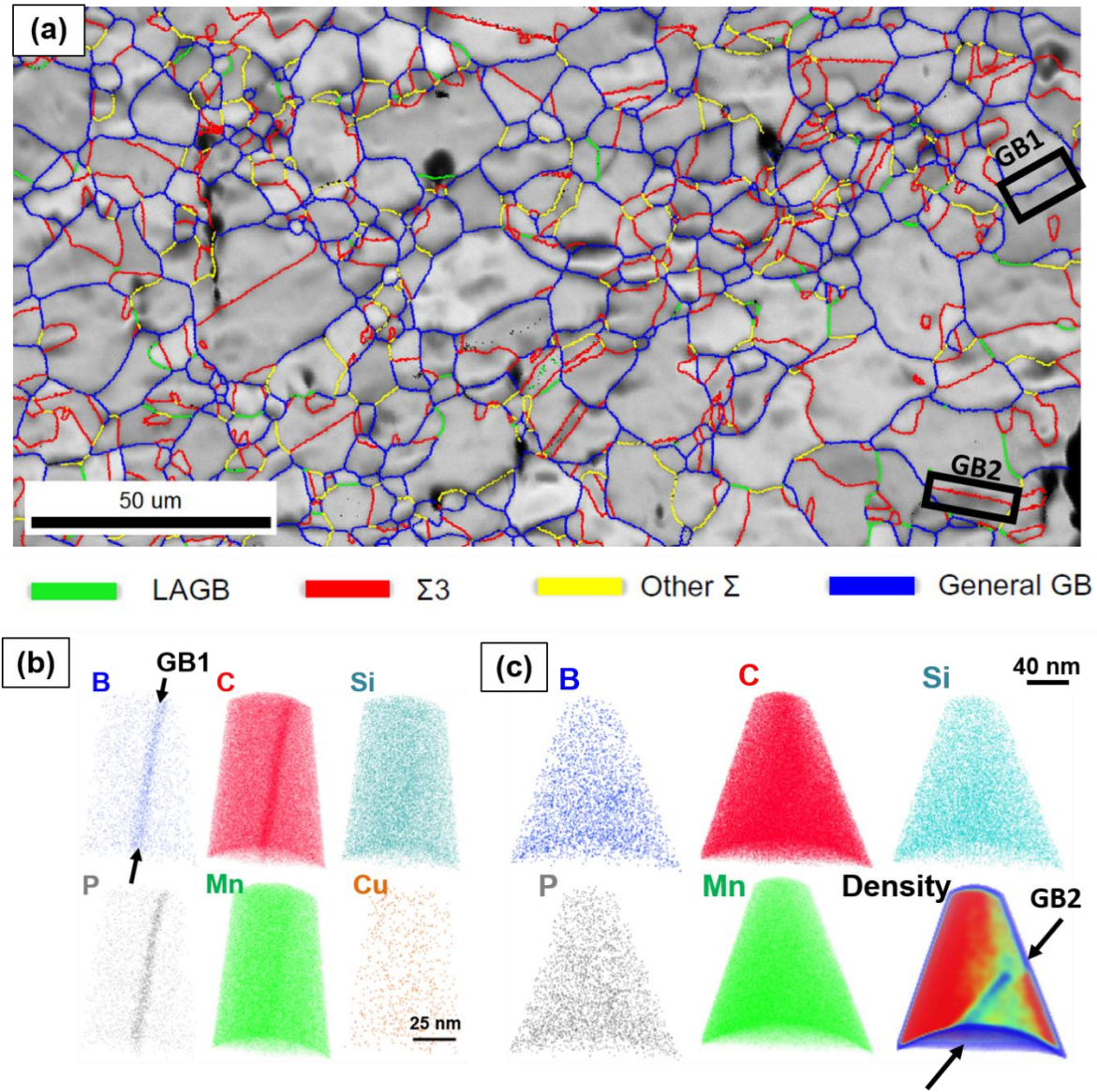

**Fig. 3** Correlative EBSD-APT analysis for Fe–28Mn–0.3C (wt.%) TWIP steel: **(a)** EBSD-based IQ+GB map showing the two GBs: GB1 and GB2 (enclosed in black outlined rectangular boxes) analysed using APT, APT-based 3D elemental maps for different elements in **(b)** GB1 and **(c)** GB2. In part **(a),** LAGB abbreviates for low-angle grain boundary and general GB refers to RHAGB. To the bottom right of part **(c)**, Density map (of Fe) shows the position of GB2 [115].

Araki et al. [116] have used the aforementioned correlative methodology coupled with nanoindentation technique to develop a systematic correlation between the critical shear stress for dislocation emission from GB and the concentration of C at (a) grain interior and (b) RHAGB in Fe-50C (ppm) ferritic steel. Based on this work, it was observed that: (i) there is a high frequency of "pop-ins" (in load-displacement curves) at RHAGB as compared to the grain interiors which suggests that dislocation nucleation and multiplication at RHAGB occurs much more easily as compared to that within grain interiors and (ii) pinning of dislocations by C atoms segregated at RHAGB leads to a high critical shear stress for emission of dislocations from RHAGB, resulting in GB strengthening [116]. Moreover, the critical shear stress required for dislocation emission from RHAGB, was quantified using Hertz contact theory considering elastic contact between the nanoindenter and the sample surface, before the onset of "pop-in events" in nanoindentation-based load-displacement curves [116]. **Table. 1** summarizes the existing literatures where TEM-APT and EBSD-APT based correlative approach have been used to characterise GB segregation in steels.

**Table. 1** A summary of reports based on correlative approach towards characterising GB segregation in steels (till date).

| Correlative approach | Steel composition (or grade) | Segregating solute species | Type of GB | Reference |
|---|---|---|---|---|
| TEM-APT | Fe–4.40C–0.30Mn–0.39Si–0.21Cr (at. %) | C | Coherent $\Sigma 5$, coherent and incoherent $\Sigma 3$ | [86] |
|  | AISI 316 | Mo, Si, Cr | RHAGB | [110] |
|  | Fe-1.62Mn-0.18Si (wt.%) | C, P | RHAGB | [111] |
| EBSD-APT | Fe-9Mn-0.05C (wt.%) | Mn | RHAGB | [112] |
|  | Fe-12Mn-3Al-0.05C (wt.%) | C, Mn | RHAGB, IB, triple junctions | [113] |
|  | Fe–0.2C–3Mn–2Si (wt.%) | C | RHAGB | [114] |
|  | Fe–28Mn–0.3C (wt.%) | B, C, P, Si | coherent $\Sigma 3$ | [115] |
|  | Fe–22Mn–0.6C (wt.%) | B, C, P, Si | $\Sigma 3$ deformation twins | [115] |
|  | Fe-50C (ppm) | C | RHAGB | [116] |

## 4. Towards utilising GB segregation: Examples
### 4.1 Alloy design

In the context of steel design, one of the areas where an understanding of GB segregation may be utilised is to stabilise nano-sized grains by reducing the overall GB energy through solute decoration at GBs. For instance, discontinuous grain growth (due to high GB energy) is a common phenomenon which has been reported in a number of materials (especially steels). In that context, GB segregation helps in two possible ways. Firstly, it reduces the GB energy and subsequently, the capillary force associated with the competitive growth of two adjoining grains. Secondly, choosing adequate solutes (for decoration at GBs) leads to an enhancement in GB cohesive strength. In the context of Fe–13.6 Cr–0.44 C (wt.%) martensitic steel, Yuan et al. [117] have reported that the ultimate tensile strength and total elongation are enhanced (by ~233.33% and 53% respectively) by the segregation of C at martensite GBs. This was attributed to the reversion from martensite to austenite phase promoted by the segregation of C at martensitic GBs [117]. Li et al. [118] have established a direct correlation between the extent of tensile deformation with the concentration of segregated C at ferrite RHAGBs in Fe–3.66C–0.48Mn–0.39Si–0.01P–0.01S (at.%) pearlitic steel. In addition, a similar correlation for the case of C segregated at ferrite sub-grain (or, low angle) boundaries has been reported in hypereutectoid Fe–4.40C–0.30Mn–0.39Si–0.21Cr–0.003Cu–0.01P–0.01S (at. %) pearlitic steel [119], [120].

The other area where an understanding of GB segregation may be utilized is to design steels with resistance to H embrittlement (HE). HE is one of the most common problems associated with automotive grade steels [121]–[129]. Being the smallest atom (with atomic radius ~0.037 nm), H easily undergoes diffusion into materials (especially steels) and leads to catastrophic and unprecedented failure of engineering components during service [122]. In addition, the extremely high atomic mobility of H (due to the small atomic radius) makes it experimentally challenging to map its exact position in a material. As a result, a number of existing strategies (devised for steels such as addition of carbides) are extremely challenging to validate. Recently, Chen et al. [124] have devised a deuterium charging based cryo-transfer protocol using characterisation techniques such as TEM, TKD and cryo-APT to identify the position of H atoms segregated at different GBs in Fe-0.23C-0.92Mn-0.24Si-0.049Nb (wt.%) steel. Two types of microstructures: (i) fully ferritic and (ii) fully martensitic were analysed using this approach [124]. Deuterium was observed to segregate at incoherent interfaces between: (i) NbC and ferrite (in the fully ferritic microstructure) and (ii) NbC and martensite laths (in the fully martensitic microstructure) [124]. This was the first experimental observation on the trapping of H atoms by a carbide. In addition, C was observed to segregate strongly at low-angle lath boundaries for the case of the fully martensitic microstructure [124]. For the case of the fully ferritic microstructure, C was observed to be strongly segregated at ferrite RHAGBs [124]. Besides, the GB segregation tendency for C atoms was reported to be influenced by GB misorientation angle in ferrite RHAGBs [124]. In both microstructures, segregation of both C and H atoms (at different GBs) were simultaneously observed [124]. However, trapping of H at GBs was reported to occur irrespective of the strong affinity of C towards H atom [124].

**4.2 Stress and segregation-induced phase transformation at a GB**

Raabe et al. [2] have highlighted that solute decoration (at GBs) coupled with local elastic stresses may be utilised to promote localised phase transformation at GBs. This is commonly observed in the context of martensite to austenite reversion at martensite GBs. The reversion phenomenon is influenced by the energy of martensite GBs [130]. Besides, the transformed region (at GBs) has been reported to facilitate further phase transformation by accommodating localised elastic stresses which leads to a localised transformation induced plasticity (TRIP) effect at the transformed austenite GBs [120], [130]. The martensite GB energy is largely dependent on the morphology of the martensite phase [49]. For instance, a number of different martensite GBs has been reported in literature which commonly include lath, needle and packet boundaries [2], [119]. Among all these boundaries, lath boundaries are associated with the minimum GB energy [2], [131]–[134]. The aforementioned reversion has been reported as a highly effective methodology to prevent intergranular crack propagation along martensite GBs. However, for the occurrence of this reversion, a number of criterion has been reported by the existing literatures [135]–[138]. First, solute species with a high $\beta_i$ must be chosen [138]. Second, the solute species must prefer to undergo preferential segregation at martensite GBs [49]. Third, the solute species must have a higher tendency towards GB segregation as compared to that for precipitate formation (for instance, the formation of carbides by a number of transition elements such as Ti, V, Cr, Nb etc.) [49]. Fourth, the segregated solute species must reduce the martensite to austenite transformation temperature [135], [136]. Finally, austenite nucleation at GBs must be associated with localised elastic stresses [137] and promote strengthening at GBs [2], [5].

**4.3 Influence of GB segregation on GB cohesive strength**

From a historical perspective, Seah [48] have investigated adsorption induced interface decohesion in commercially pure α-Fe and initially considered the differences between two types of boundary separation as influenced by solute adsorption: (i) quasi-equilibrium separation with the chemical potential of solute maintained uniform throughout the system and

(ii) "rapid" separation in a manner that the excess amount of solute initially residing in the boundary remains attached to the created free surfaces with no solute exchange taking place between bulk phases [1]. Based on it, a model to theoretically calculate GB cohesive strength using sublimation enthalpies and atomic sizes has been presented [139]. Besides, it has been reported that during fracture, dislocations are emitted across the region of the crack tip and contribute to the plastic work in the fracture process [140]. If the GB strengthening tendency is reduced, failure can occur at reduced stress levels and hence the overall plastic work involved also gets reduced [1]. Before 1979, Thermochemical model was used for the theoretical study of GB cohesion which assumed that GB separates without redistribution of the solute atoms during fracture [1]. However, Seah [48] presented a Pair Bonding or quasi-chemical model to provide a simplified numerical evaluation of the behaviour of embrittlement in α-Fe. The Pair Bonding theory states that the actual energy required to break the bonds across the GB may be simply determined by counting the number of dangling bonds per unit area and adding their energies [48], [141]–[143]. The model predicted that in a Fe matrix, Bi, S, Sb, Se, Sn and Te will be highly embrittling and will lead to intergranular separation under application of mechanical stress, followed in order of reducing effect by P, As, Ge, Si and Cu [2], [48]. In increasing order of their remedial effect on embrittlement in α-Fe are N, B and C which can prevent intergranular fracture and thus, improve the overall ductility, by maintaining the GB unruptured [1], [2], [48].

Gibson and Schuh [142], [144] have proposed Bond-breaking model for calculating the change in GB cohesive energy as a function of equilibrium segregation of solute species in α-Fe and developed GB segregation and associated cohesion maps for determining whether a particular combination of solute and solvent species will lead to GB embrittlement/strengthening. Based on this model, the energetic barrier to intergranular cleavage may be termed as GB cohesive energy ($E_{GBCE}$) and is given as: [144]

$$E_{GBCE} = 2\gamma_s - \gamma_{GB} \qquad (6)$$

where, $\gamma_s$ is the surface energy. Both these energies are largely influenced by the alloy composition and temperature. Hirth and Rice [145] have presented a mathematical framework for representing $E_{GBCE}$ as a function of $\Gamma_i$ for solute species segregated at GBs. This thermodynamic framework is based on the assumption that there is no segregation of solute atoms at GBs during intergranular cleavage [145], [146]. The model is given as: [145]

$$E_{GBCE}(\Gamma) = E_{GBCE}(\Gamma = 0) - \int_0^\Gamma \left(\mu_{GB}(\Gamma) - \mu_S\left(\frac{\Gamma}{2}\right)\right) d\Gamma \qquad (7)$$

where, $\mu_{GB}$ and $\mu_S$ are the chemical potentials at GB and surface respectively [145]. The integral in equation (7) describes the change in $E_{GBCE}$ of a pure metal upon solute addition [145], [147]–[151]. However, there is no emphasis on the interaction between multiple solute species for multi-element segregation at GBs. Besides, there is no term to account for GB energy in equation (7).

GB cohesive strength is highly influenced by the extent of solute segregation at GBs and free surfaces (FS) [140]. The preference of a solute to segregate in either GB or FS depends on the magnitude of binding (or segregation) energy at GB and FS [140], [146]. For instance, when GB segregation energy is higher than FS segregation energy, the solute atoms reduce GB cohesive strength and hence act as potential GB embrittlers [140]. In cases where the FS segregation energy is higher than that of GB segregation energy, the solute atoms enhance the GB cohesive strength [140]. The thermodynamic basis for such an inference is based on the Rice-Wang Model [134]. Rajagopalan et al. [140] have theoretically investigated the influence of substitutional elements (such as V and P) on the GB cohesive strength of symmetrical tilt $\Sigma5(210)$ ($\theta=53.13°$) SHAGBs in α-Fe. It was shown that P reduces the GB cohesive strength of $\Sigma5$ GB whereas V increases the GB cohesive strength of the same [140]. **Table. 2** shows the proposed theoretical models (till date) for determining GB cohesive energy as a function of solute decoration at RHAGB in α-Fe.

**Table 2.** Theoretical models on determining GB cohesive energy as a function of solute decoration at RHAGB in α-Fe.

| Proposed model | Main feature (or assumption) | Reference |
|---|---|---|
| Thermochemical | GB cleavage occurs without solute redistribution. | [47] |
| Pair Bonding or quasi-chemical | Energy required to break the bonds across a GB may be determined by calculating the number of dangling bonds per unit area and adding their energies. | [48] |
| Hirth-Rice | Constant chemical potential of the segregating species is maintained during GB cleavage. | [145] |
| Rice-Wang | GB cleavage occurs without any local stress concentration. | [146] |
| Bond-breaking | Solute (segregated at a GB) does not undergo diffusion during GB cleavage | [147] |

Silva et al. [152] have introduced a model for correlating GB segregation with intergranular embrittlement and GB phase nucleation by coupling the thermodynamics of GB segregation in solid solutions with GB character. The proposed model was experimentally validated the model with Atom Probe Tomography (APT)-based investigations in Fe-Mn binary alloys [152]. It was shown that an increase in solute concentration at RHAGBs (BCC RHAGBs in the context of Fe-Mn alloys) leads to an increase in $\Delta H_i$ and triggers GB embrittlement at room temperature [152]. However, once austenitic transformation occurs, there is a drastic decrease in solute decoration at RHAGBs leading to an increase in GB cohesive strength [152]. The aforementioned model has also been extended to Fe-9Mn (wt.%) based steels [19], [113], [153], [154]. Yoo et al. [155] have reported that addition of 0.15 at.% Mo leads to an improvement in HE resistance in a 32MnB5 hot-stamping steel. This was attributed to the reduction of H diffusivity by Mo [155]. Besides, Mo was also reported to enhance GB cohesive strength and reduce strain localisation along austenite RHAGBs leading to a transition in crack propagation from intergranular to transgranular mode [155]. In the context of Fe-1.62Mn-0.18Si (wt.%) ferritic steel, segregation of P (at ferrite RHAGBs) has been reported to drastically reduce GB cohesive strength [111]. On the other hand, the presence of C (at RHAGBs) has been observed to promote strengthening at GBs [111].

## 5. Experimental challenges with GB research from the viewpoint of solute decoration

A major challenge associated with GB research is to address all five macroscopic and three microscopic DOFs (associated with GBs). Although 3D EBSD technique (proposed in Ref. [156]) addresses all five DOFs (in a GB), however, experimentally, it has still not been possible to address the three microscopic DOFs. Besides, 3D EBSD technique is highly time-consuming and requires an additional FIB setup in an SEM. Hence, unlike conventional 2D-EBSD technique, this characterisation cannot be performed in a conventional SEM. The inability of addressing the eight DOFs leads to limitations in terms of correlating the GB structure and solute decoration with their overall influence on the mechanical performance of the material. However, when utilising GB segregation information to reduce the overall grain size, the only information required is the extent of GB segregation. In this context, information about GB plane may be neglected and hence, a full 5D GB analysis is not required [2]. In addition, 3D-EBSD uses serial sectioning and hence, may be regarded as a destructive technique [156], [157]. In other words, it is not possible to devise a correlative 3D-EBSD-APT methodology for the purpose of obtaining both structural and chemical information from the same region in a microstructure as because the analysed region is already lost during 3D-EBSD mapping and hence, not available for APT analysis [157]. Nevertheless, the understanding on how the concentration of segregating solute species and dislocation density (at pile-ups) near GBs, influences GB cohesive strength still remains limited. This may be attributed to the numerous experimental challenges associated with determining the interaction between a GB and the decorating solute species. The correlative methodology of microstructural characterisation (discussed in section **3.2.2**) coupled with appropriate theoretical analyses may be used as a potential tool to address the above challenge.

## 6. Conclusions and outlook

Solute decoration at internal interfaces leads to a number of phenomenon which commonly include strain-induced phase transformation (at the internal interfaces), bulk phase formation, phase reversion and intergranular embrittlement. From a metallurgical viewpoint, interfacial decoration may be utilised for both structural and chemical manipulation of internal interfaces. The primary aim behind the aforementioned manipulation is to enhance the overall mechanical performance of steels. Nevertheless, this is also an avenue where the methodology of correlative microscopy (coupled with appropriate theoretical analyses) may be utilised as a potential tool to design high-performance steels.

### Acknowledgement

The author hereby declares that the review article is solely authored by him. No external assistance of any kind has been received for the article.